\documentclass{article}
\usepackage[utf8]{inputenc}
\usepackage[a4paper,left=3cm,right=3cm,top=3cm,bottom=3cm]{geometry}
\usepackage{amsmath}
\usepackage{amssymb}
\usepackage{graphicx}
\usepackage{enumitem}
\usepackage{esint}
\usepackage{hyperref}
\usepackage{tikz}
\usepackage{cite}
\hypersetup{
  colorlinks   = true,    % Colours links instead of ugly boxes
  urlcolor     = blue,    % Colour for external hyperlinks
  linkcolor    = blue,    % Colour of internal links
  citecolor    = black     % Colour of citations
}

\newcommand{\KaTie}{{\sc Ka\hspace{-0.2ex}Tie}}
\newcommand{\Pythia}{{\sc Pythia}}

\usepackage{color}
\usepackage[makeroom]{cancel}
\usepackage[normalem]{ulem}
\definecolor{pkcolor}{rgb}{0,0.1,0.7}

\newcommand\pkout{\marginpar{\color{pkcolor}$\clubsuit$}\bgroup\markoverwith{\color{pkcolor}{\rule[0.4ex]{2pt}{0.8pt}}}\ULon}

\definecolor{kkcolor}{rgb}{1,0,0}

\newcommand\kkout{\marginpar{\color{kkcolor}$\int$}\bgroup\markoverwith{\color{kkcolor}{\rule[0.4ex]{2pt}{0.8pt}}}\ULon}

\definecolor{hkcolor}{rgb}{0.7,0.0,0.0}

\newcommand\hkout{\marginpar{\color{hkcolor}$\clubsuit$}\bgroup\markoverwith{\color{hkcolor}{\rule[0.4ex]{2pt}{0.8pt}}}\ULon}

\definecolor{macolor}{rgb}{0.5,0.2,0.7}

\newcommand\maout{\marginpar{\color{macolor}$\int$}\bgroup\markoverwith{\color{macolor}{\rule[0.4ex]{2pt}{0.8pt}}}\ULon}

\definecolor{pvmcolor}{rgb}{1.0,0.5,0.5}

\newcommand\pvmout{\marginpar{\color{pvmcolor}$\int$}\bgroup\markoverwith{\color{pvmcolor}{\rule[0.4ex]{2pt}{0.8pt}}}\ULon}

\title{Dijet azimuthal correlations in p-p and p-Pb collisions at forward LHC calorimeters}
\author{
M. Abdullah Al-Mashad$\,\,\,^a$ A. van Hameren$\,\,\,^c$ H. Kakkad$\,\,\,^d$ P. Kotko$\,\,\,^d$ \\
K. Kutak$\,\,\,^{c,e}$\,\,\,P. van Mechelen$\,\,\,^b$ S. Sapeta$\,\,\,^c$ \\
\\ \\
$^a$ {\it Fayoum University, Center for High Energy Physics (CHEP-FU),} \\
     {\it Department of Physics, Faculty of Science,} \\
     {\it  Fayoum 63514, Egypt } \\\\
$^b$ {\it Antwerp University, Particle Physics group} \\
     {\it  Groenenborgerlaan 171, 2020 Antwerpen, Belgium } \\\\
$^c$ {\it Institute of Nuclear Physics, Polish Academy of Sciences} \\
     {\it  Radzikowskiego 152, 31-342 Krak\'ow, Poland } \\ \\
$^d$ {\it AGH University Of Science and Technology, }\\
     {\it Faculty of Physics and Applied Computer Science,} \\ 
    {\it al. Mickiewicza 30, 30-059 Krak\'ow, Poland} \\ \\
$^e$ {\it Brookhaven National Laboratory, }\\
     {\it Physics Department, Bldg. 510A,} \\ 
    {\it 20 Pennsylvania Street, 30-059 Upton, NY 11973 USA} 
    }

\date{}

\begin{document}
%=======================================================================
\maketitle

%--------------- preprint numbers ---------------------
\vspace{-44em}
\begin{flushright}
  IFJPAN-IV-2022-17\\
\end{flushright}
\vspace{40em}
%-----------------------------------------------------------------

\begin{abstract}
    We present a state-of-the-art computation for the production of forward dijets in proton-proton and
    proton-lead collisions at the LHC, in rapidity domains covered by the ATLAS calorimeter and
    the planned FoCal extension of the ALICE detector. We use the small-$x$ improved TMD (ITMD) formalism,
    together with collinearly improved TMD gluon distributions and full $b$-space Sudakov resummation,
    and discuss nonperturbative corrections due to hadronization and showers using the \Pythia\ event generator. We observe that forward dijets in proton-nucleus collisions at moderately low $p_T$ are excellent probes of saturation effects, as the Sudakov resummation does not alter the suppression of the cross section.
\end{abstract}

%-----------------------------------------------------------------------
\section{Introduction}
\label{sec:Intro}

One of the current experimental challenges in Quantum Chromodynamics (QCD) are searches for clean signals of gluon saturation, i.e.\@ a signature of gluon recombination in a dense nuclear system. Gluon saturation has been predicted from QCD long time ago \cite{Gribov:1984tu} and  has been systematically studied over the years, in particular using the Color Glass Condensate (CGC) effective theory (see e.g.\@ \cite{Gelis:2010nm}). Although there is no doubt that the growth of gluon distributions has to be tamed at some point due to the unitarity of a scattering matrix, and while there are strong hints for occurrence of saturation in data \cite{Golec-Biernat:1998zce,Stasto:2000er,Albacete:2010pg,Dusling:2007gi,Ducloue:2015gfa,ArroyoGarcia:2019cfl,Benic:2022ixp,Armesto:2022mxy} (see \cite{Morreale:2021pnn} for a review), there is no complete consensus on how the very small $x$ limit is reached. Moreover, it is expected to see the onset of Balitsky-Fadin-Kuraev-Lipatov (BFKL) dynamics \cite{Kuraev:1977fs, Balitsky:1978ic}, even before  saturation dynamics. One example is Mueller-Navalet jet production \cite{Colferai:2010wu,Ducloue:2013wmi,Ducloue:2013bva} in the $k_T$ factorization formalism\cite{Catani:1990eg}  (see also \cite{Celiberto:2020wpk} for recent developments) \cite{Ball:2017otu} for  BFKL resummation in collinear factorization or central-forward inclusive jets at the LHC \cite{Deak:2009xt,Deak:2010gk,Deak:2011ga}. 

However, there is an important difference between  BFKL and saturation physics, which could potentially allow for saturation to be seen more directly. Saturation phenomena manifest themselves through the high energy evolution equations, similar to BFKL, but nonlinear (the Balitsky-Kovchegov (BK) equation \cite{Balitsky:1995ub,Kovchegov:1999yj} and the  B-JIMWLK equations \cite{Balitsky:1995ub,
JalilianMarian:1997jx,JalilianMarian:1997gr,JalilianMarian:1997dw,Kovner:2000pt,Kovner:1999bj,Weigert:2000gi,Iancu:2000hn}. The strength of the nonlinearity taming the growth of gluon distributions strongly depends on the target size -- for large systems with $A$ nuclei it is expected to be enhanced by roughly $A^{1/3}$.
Therefore, comparing observables computable within the high energy QCD limit for a proton and for large nuclear targets is potentially the best way to find evidence for saturation. 
Such dependence of the cross section for production of forward $\pi^0$ in p+A was recently reported in \cite{STAR:2021fgw}, providing strong signs of saturation.
It is important to mention that there might be other mechanisms giving suppression of nuclear parton distribution functions (PDFs), notably the so-called leading twist nuclear shadowing \cite{Frankfurt:2011cs} which is used within  collinear factorization. However, at present its connection to saturation is unclear, although one has to keep in mind that the saturation for dijet production is 
also the leading power effect.

In our work we are interested in dijet production as a probe of saturation (see \cite{Kharzeev:2004bw,Marquet:2007vb,Kutak:2012rf,vanHameren:2014lna,Kutak:2014wga} for earlier works on this subject) in hadro-production. We thus require the final state partons to have rather large transverse momenta $P_T$. Naturally, the scale set by the jets is  larger than the saturation scale $Q_s$, but not asymptotically larger, so that the saturation effects are not neglected. Such limit is well defined within the CGC theory, and is precisely the leading power limit $k_T/P_T \ll 1$, where $k_T$ is the dijet imbalance \cite{Dominguez:2011wm}. In our computations we go beyond the leading power, by including the kinematic twists -- such approach gives more precise predictions for the dijet correlation spectra. The adequate formalism is known as the small-$x$ improved Transverse Momentum Dependent (ITMD) factorization \cite{Kotko:2015ura,Altinoluk:2019fui} (for further developments of both the ITMD and the leading power limit see \cite{vanHameren:2016ftb,Marquet:2016cgx,Marquet:2017xwy,Bury:2018kvg,Altinoluk:2019wyu,Altinoluk:2018byz,Altinoluk:2020qet,Fujii:2020bkl,Altinoluk:2021ygv,Boussarie:2021ybe,Taels:2022tza}). 

For dijet imbalance observables it is necessary to perform a suitable resummation of the Sudakov logs. This can be done in at least two ways. First method relies on including the Sudakov form factor as a source of the hard scale evolution, similar to what is being done in parton shower algorithms. Such approach has been used for instance in \cite{vanHameren:2014ala,vanHameren:2015uia,Kutak:2014wga,vanHameren:2019ysa}. Another approach relies on the soft gluon resummation technique in $b$-space \cite{Mueller:2013wwa, Mueller:2012uf}, which in general provides resummation beyond simple double Sudakov logs (see e.g. \cite{Zheng:2014vka,Stasto:2018rci,Benic:2022ixp}. In the present work, we shall apply the full $b$-space resummation approach, as a current state-of-the-art result.

Forward jets have been already measured at LHC, with inconclusive result regarding the saturation signal. For example, the CMS-CASTOR calorimeter \cite{CMS:2020ldm} measured single inclusive jets \cite{CMS:2018yhi} in proton-lead collision, but the lack of the proton-proton study makes it very difficult to assess if saturation is present. This is mainly due to the fact that at present all saturation-based calculations are parton-level and thus the comparison with data is burdened with large uncertainties \cite{Kutak:2017ncj,Bury:2017xwd,Mantysaari:2019nnt}
. Further, the ATLAS collaboration measured forward-forward and forward-central dijets \cite{ATLAS:2019jgo} for both proton-proton and proton-lead, but no cross section measurement has been done, and thus no nuclear modification ratio was provided. The visible nuclear broadening has been claimed to be negligible within the error bars, despite being consistent with saturation and Sudakov resummation \cite{vanHameren:2019ysa}. Finally, the CMS collaboration recently measured exclusive dijet production \cite{CMS:2022lbi} in ultra-peripheral collisions, where, again, only the photon-lead sample is studied, without a photon-proton reference. Interestingly, a comparison with a Monte Carlo describing the photoproduction on proton targets seems to imply strong nuclear broadening. 

In the present work we provide predictions for a potential new study of forward dijets with ATLAS FCal kinematics, as well as for the planned FoCal upgrade of ALICE \cite{ALICECollaboration:2719928}, assuming that both proton-proton and proton-lead cross sections will be measured. Our paper is organized as follows. In the next Section we briefly review the ITMD framework and modify it accordingly to include the Sudakov resummation. Next, in Section~\ref{sec:Results} we specify our kinematic cuts in detail and present our results.  We delegate the discussion of the results to Section~\ref{sec:Summary}.

%-----------------------------------------------------------------------
\section{Small-$x$ Improved TMD Factorization}
\label{sec:ITMD}

The ITMD factorization formula for the production of two jets with momenta $p_1$ and $p_2$, and rapidities $y_1$ and $y_2$, reads
\begin{equation}
\frac{d\sigma^{\mathrm{pA}\rightarrow j_1j_2+X}}{d^{2}P_{T}d^{2}k_{T}dy_{1}dy_{2}}
=
%\frac{1}{(x_{\mathrm{p}} x_{\mathrm{A}} s)^{2}}
\sum_{a,c,d} x_{\mathrm{p}} f_{a/\mathrm{p}}\left(x_{\mathrm{p}},\mu\right) 
\sum_{i=1}^{2}\mathcal{K}_{ag^*\to cd}^{\left(i\right)}\left(P_T,k_T;\mu\right)
\Phi_{ag\rightarrow cd}^{\left(i\right)}\left(x_{\mathrm{A}},k_T \right)\,,
\label{eq:itmd}
\end{equation}  
 where for jets with transverse momenta $\vec{p}_{T1}$ and $\vec{p}_{T2}$ we defined $\vec{P}_T=\vec{p}_{T1}-\vec{p}_{T2}$ and the dijet transverse momentum 
imbalance $\vec{k}_T=\vec{p}_{T1}+\vec{p}_{T2}$. The longitudinal fractions of partons extracted from proton and nucleus are, respectively,  $x_{\mathrm{p}}$ and $x_{A}$, with a restriction that $x_A\ll x_{\mathrm{p}}$.
Furthermore, $f_{a/\mathrm{p}}$ are collinear PDFs, $\mathcal{K}_{ag^*\to cd}$ are off-shell gauge invariant hard factors and
$\Phi_{ag\rightarrow cd}^{\left(i\right)}$ are the TMD gluon distributions that correspond to distinct color flows for each partonic channel. The hard factors and the TMD gluon distributions were computed in
\cite{Kotko:2015ura}.

 The resummation of the Sudakov logarithms is performed following the perturbative calculation presented in \cite{Mueller:2013wwa},
where the calculations have been done in the impact parameter space (here, the impact parameter $b_T$ is the Fourier conjugate to the gluon $k_T$).
The derivation has been done in the back-to-back regime, that is to leading power. 
Since the Sudakov factors are negligible for $k_T\sim P_T$, it can be straightforwardly extended to the ITMD formula \eqref{eq:itmd}. 
\begin{multline}
\frac{d\sigma^{\mathrm{pA}\rightarrow j_1j_2+X}}{d^{2}P_{T}d^{2}k_{T}dy_{1}dy_{2}}
=
\sum_{a,c,d} x_{\mathrm{p}} 
\sum_{i=1}^{2}\mathcal{K}_{ag^*\to cd}^{\left(i\right)}\left(P_T,k_T;\mu\right) \\
\times\int db_T b_T J_0(b_T k_T) 
f_{a/\mathrm{p}}\left(x_{\mathrm{p}},\mu_b\right) 
\widetilde{\Phi}_{ag\rightarrow cd}^{\left(i\right)}\left(x_{\mathrm{A}},b_T\right)
e^{-S^{ag\to cd}(\mu,b_\perp)} \,,
\label{eq:itmd_Sud}
\end{multline}
where $\widetilde{\Phi}_{ag\rightarrow cd}^{\left(i\right)}$ is the Fourier transform of the TMD gluon distributions and 
$S^{ag\to cd}$ are the Sudakov factors defined below. The scale $\mu_b$ is essentially the inverse of the impact parameter:
\begin{equation}
\mu_b=2e^{-\gamma_E}/b_*    
\end{equation}
with
\begin{equation}
\label{eq:bstar-def}
 b_* = b_T/\sqrt{1+b_T^2/b_{\rm max}^2}\,.
\end{equation}
With such a choice, the scale $\mu_b$ freezes in the limit of large $b_T$, where it takes the value $2 e^{-\gamma_E}/b_{\max} \gg \Lambda_\text{QCD}$. Following Ref.~\cite{Marquet:2019ltn}, in our calculation we shall use the value $b_{\max} = 0.5\, \text{GeV}^{-1}$.

For each channel, the Sudakov factors can be written as
\begin{equation}
  S^{ab\to cd} (\mu,b_\perp) =\sum_{i=a,b,c,d} S_p^{i} (\mu,b_\perp) +
\sum_{i=a, c, d}S^{i}_{np} (\mu,b_\perp), 
\end{equation} 
where $S_p^{i}$ and $S_{np}^{i}$ are the perturbative and
non-perturbative contributions. 
It was argued in~Ref.~\cite{Stasto:2018rci}, that the non-perturbative Sudakov should not be included for a small-$x$ parton $b$.
The perturbative Sudakov factors, including double and single logarithms, are given by~\cite{Mueller:2012uf,Mueller:2013wwa}
\begin{gather}
S_p^{qg\to qg} (\mu, b_\perp) = \int_{\mu_b^2}^{\mu^2} \frac{dq_T^2}{q_T^2} \left[
2 (C_F + C_A) \frac{\alpha_s}{2\pi} \ln \left( \frac{\mu^2}{q_T^2} \right)
- \left(\frac{3}{2}C_F +  C_A \beta_0 \right) \frac{\alpha_s}{\pi}
  \right],
  \label{eq:sudpertqg}
  \\
S_p^{gg\to gg} (\mu, b_\perp) =  \int_{\mu_b^2}^{\mu^2} \frac{dq_T^2}{q_T^2} \left[
  4 C_A \frac{\alpha_s}{2\pi} \ln \left( \frac{\mu^2}{q_T^2} \right)
- 3 C_A \beta_0 \frac{\alpha_s}{\pi} \right]\,,
  \label{eq:sudpertgg}
\end{gather}
where 
$\beta_0 = (11-2n_f/N_c)/12$. 
The $g g \to q \bar q$ channel is negligible for the kinematic domain of this
study\footnote{The single logarithm accuracy terms have been recently obtained at leading power within the small-$x$ CGC formalism for di-jet production in e-A at NLO accuracy~\cite{Caucal:2022ulg}. }.

Let us notice, that in \eqref{eq:itmd_Sud} the collinear PDF depends on the impact parameter.
This complicates the Monte Carlo implementation of the factorization approach. Therefore we investigate a choice of the 
factorization scale, which is independent on $b_T$. As argued for example in \cite{Xiao:2018zxf} this formally introduces  
a threshold-type logarithmic term. The real impact of this term in the kinematic domain under study is however difficult to judge, without
concrete computations. 
Setting $\mu_b=\mu$, the collinear PDF factorizes outside the $b$-space integral and we can define the hard scale-dependent TMD gluon distribution as 
\begin{align}
   \Phi_{ag\rightarrow cd}^{\left(i\right)}(x,k_\perp,\mu) & =
    \int d b_\perp \int dk^{\prime}_\perp \, b_\perp\, k^\prime_\perp\, 
    J_0(b_\perp\,k^\prime_\perp)\, 
    J_0(b_\perp \,k_\perp)\,
    \nonumber \\
    & \hspace{50pt}\times
    {\cal F}_{g^*/B}(x,k^\prime_\perp)\, e^{-S^{ag\to cd}(\mu,b_\perp)}\,.
    \label{eq:gluon-dff}
\end{align}
The above TMD gluon distribution can then be straightforwardly used in the factorization formula \eqref{eq:itmd}.
The hard scale $\mu$ in the dijet production process is provided by the jet transverse momentum. Specifically, in our computations we shall use the the average $p_T$ of the two leading jets.

 In order to compare the above approach to the full $b$-space resummation, we apply the following procedure, that can be relatively easily implemented in
a Monte Carlo program.
First one generates events in the simplified approach with $\mu_b=\mu$. Then, just for the generated space phase points, one calculates the following quantity:
\begin{align}
   \big(f_{a/\mathrm{p}}\otimes \Phi_{ag\rightarrow cd}^{\left(i\right)}\big)(x_{\mathrm{p}},x,k_\perp,\mu) & =
    \int d b_\perp \int dk^{\prime}_\perp \, b_\perp\, k^\prime_\perp\, 
    J_0(b_\perp\,k^\prime_\perp)\, 
    J_0(b_\perp \,k_\perp)\,
    \nonumber \\
    &
    \hspace{4pt} \times
    f_{a/\mathrm{p}}\left(x_{\mathrm{p}},\mu_b\right) 
    {\cal F}_{g^*/B}(x,k^\prime_\perp)\, e^{-S^{ag\to cd}(\mu,b_\perp)}\,.
    \label{eq:gluon-bdep}
\end{align}
Finally, in order to obtain the full $b$-space resummation, one reweighs the events with a ratio
\begin{equation}
\frac{\big(f_{a/\mathrm{p}}\otimes \Phi_{ag\rightarrow cd}^{\left(i\right)}\big)(x_{\mathrm{p}},x,k_\perp,\mu)}{f_{a/\mathrm{p}}(x_{\mathrm{p}},\mu)\,\Phi_{ag\rightarrow cd}^{\left(i\right)}(x,k_\perp,\mu)}
~.
\end{equation}
As we shall see in the next Section, both approaches give very similar results, validating the simplified approach for forward dijet production processes.

%=========================================================
%=========================================================
\section{Numerical results}
\label{sec:Results}

In this section we present our results for:
\begin{itemize}%[leftmargin=35pt]
    \item differential cross sections as a function of the azimuthal angle $\Delta\Phi$ between the leading and sub-leading jets,  both for p-p and p-Pb collisions at $\sqrt{s}=8.16\, \mathrm{TeV}$,
    \item nuclear modification ratios, necessary to quantify saturation effects, defined as:
\begin{equation}
    R_{\mathrm{p-Pb}}=\frac{\frac{d \sigma^{p+P b}}{d \mathcal{O}}}{A \frac{d \sigma^{p+p}}{d \mathcal{O}}}\,.
\end{equation}
\end{itemize}

\begin{figure} 
  \begin{minipage}[b]{0.5\linewidth}
    \centering
    \includegraphics[width=1\linewidth]{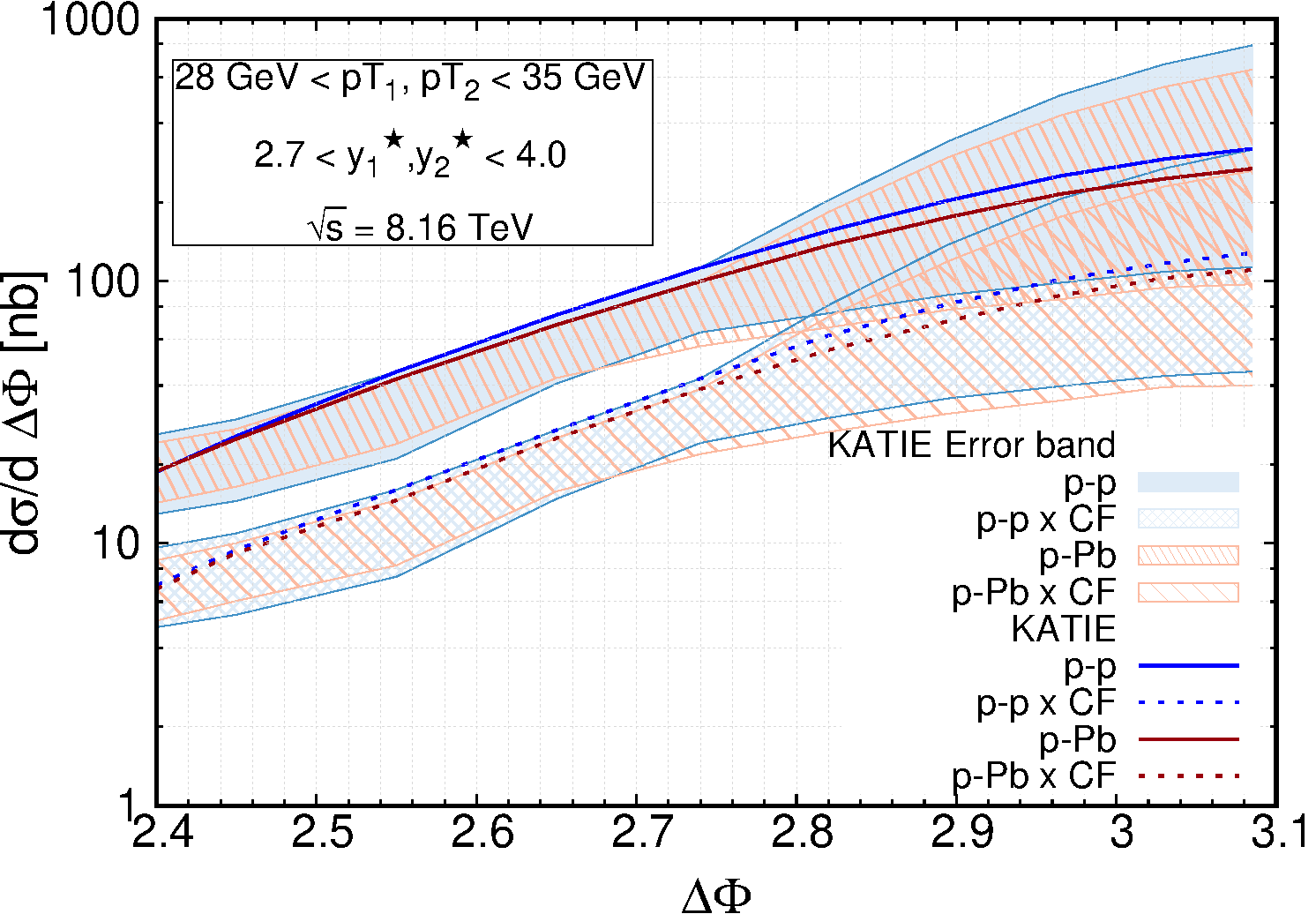} 
   %\caption{(a)} 
    \vspace{1ex}
  \end{minipage}%%
  \begin{minipage}[b]{0.5\linewidth}
    \centering
    \includegraphics[width=1\linewidth]{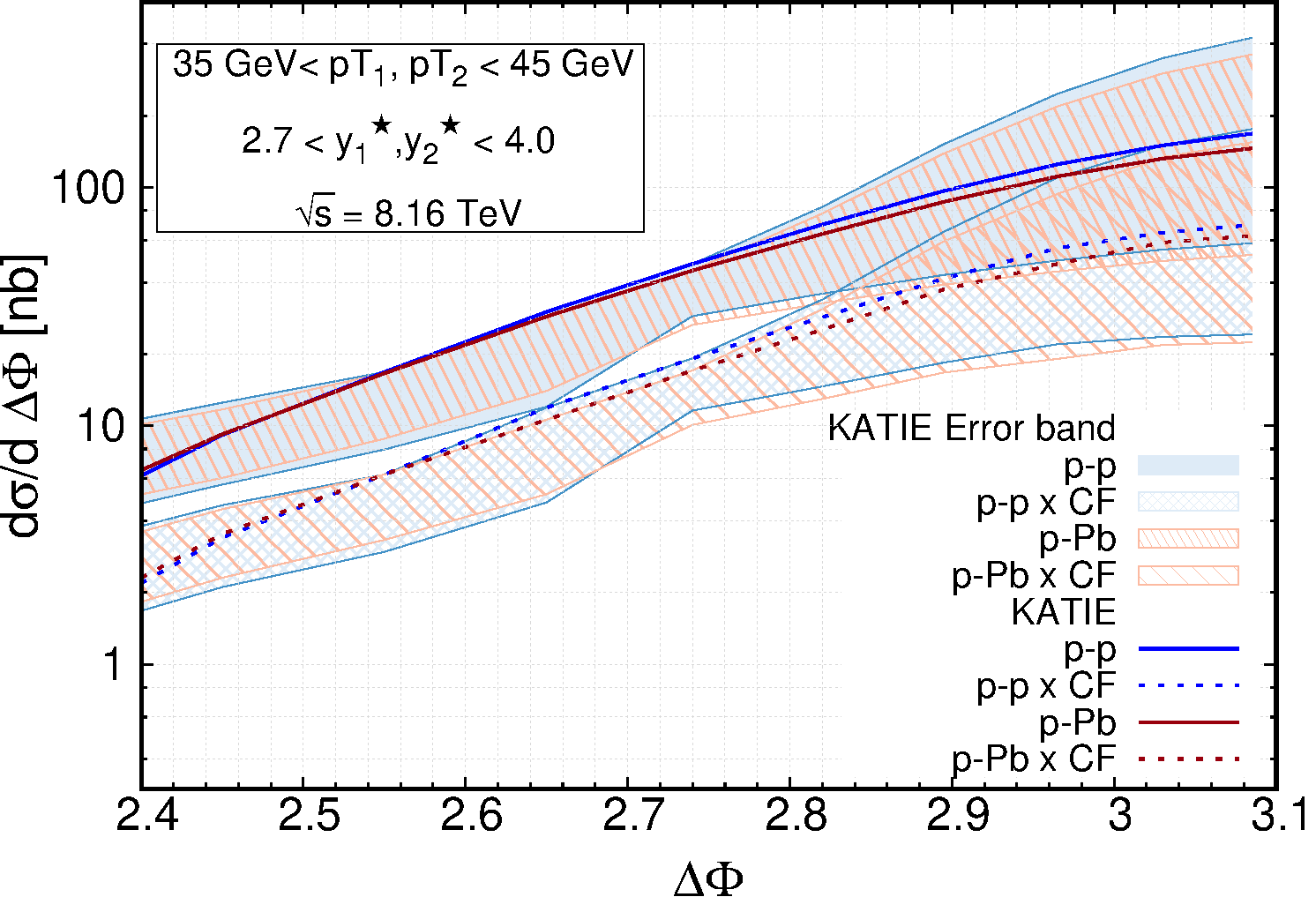} 
    %\caption{At2} 
    \vspace{1ex}
  \end{minipage} 
  \begin{minipage}[b]{0.5\linewidth}
    \centering
    \includegraphics[width=1\linewidth]{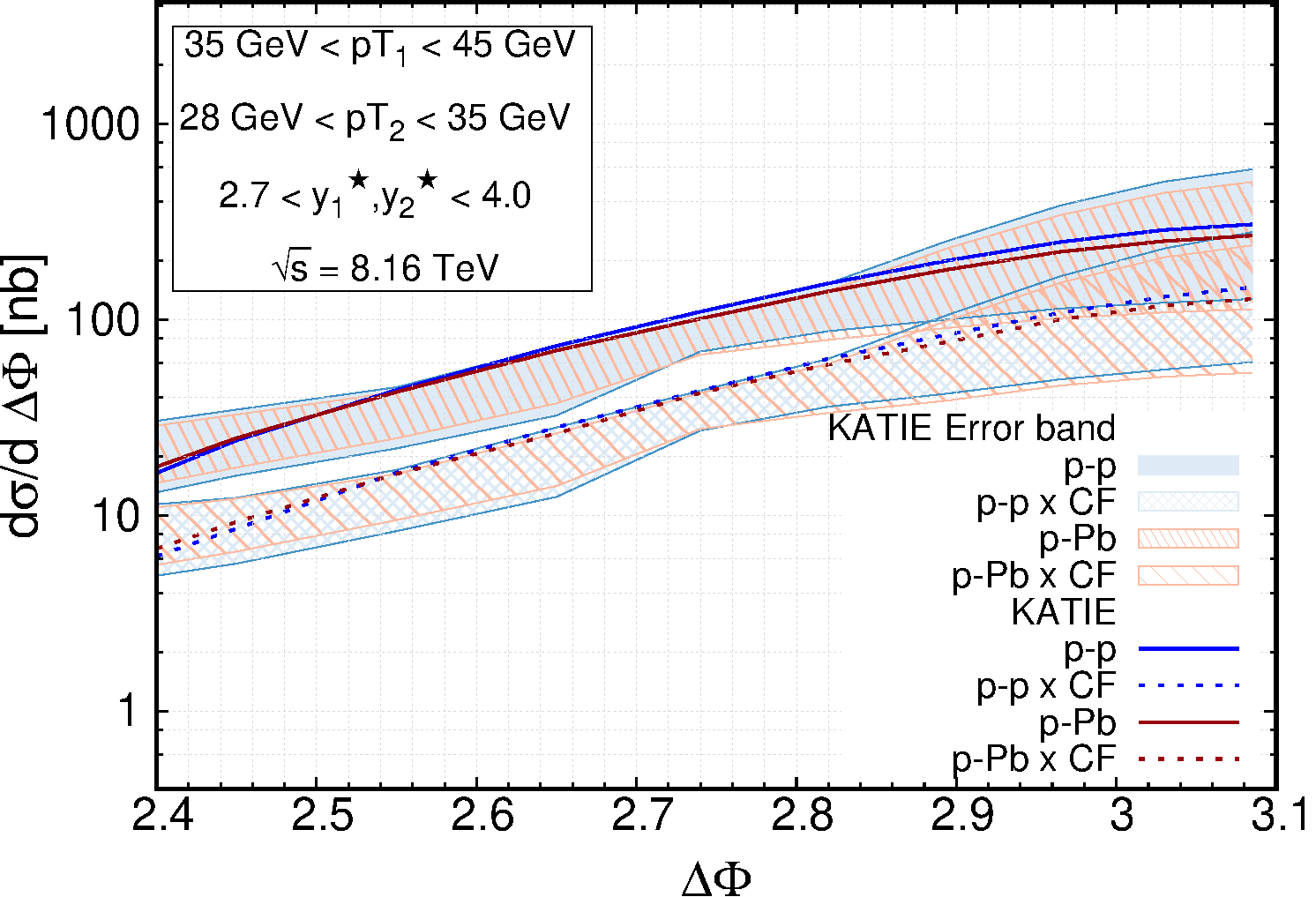} 
   % \caption{AT3} 
    \vspace{1ex}
  \end{minipage}%% 
  \begin{minipage}[b]{0.5\linewidth}
    \centering
    \includegraphics[width=1\linewidth]{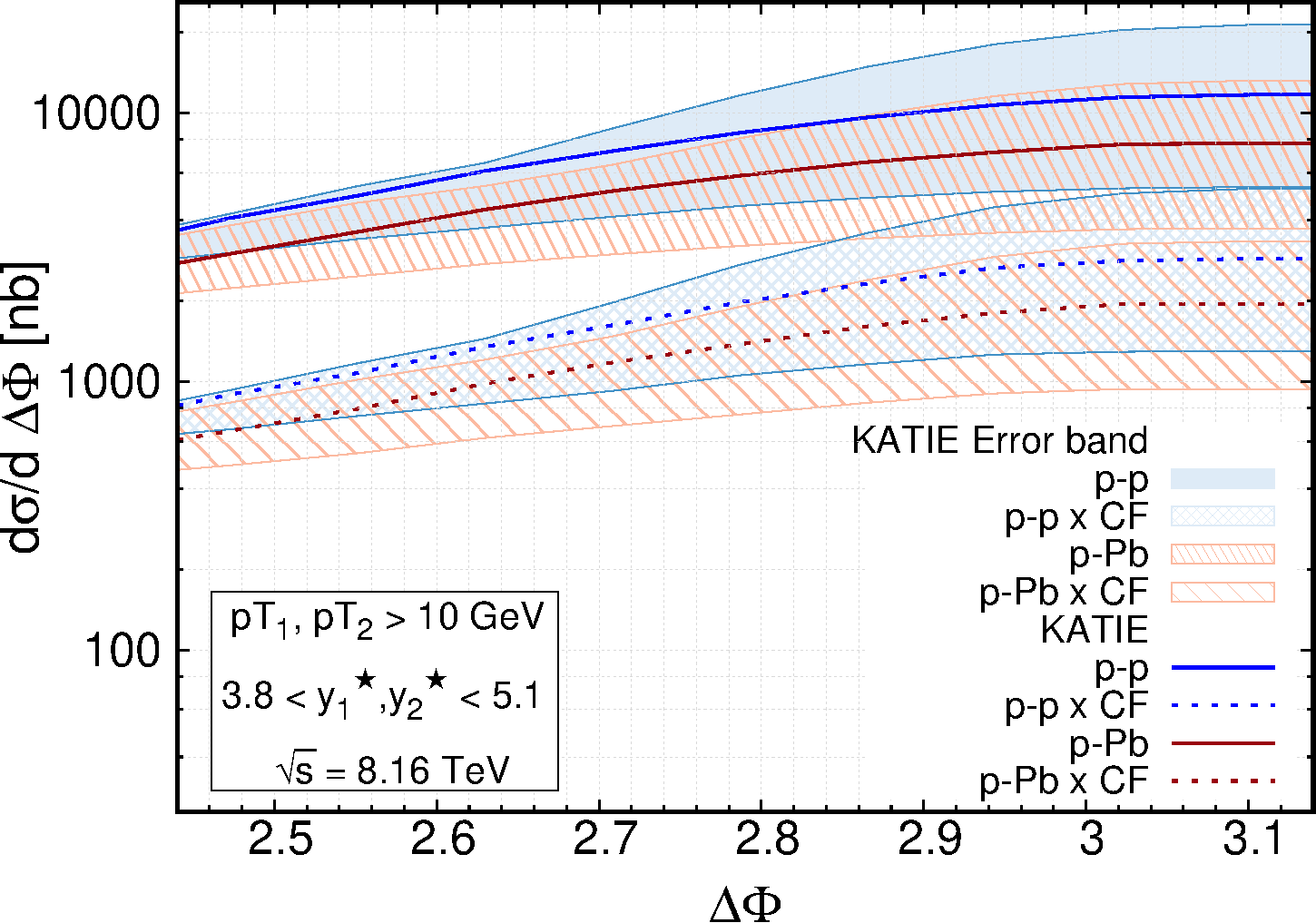} 
   % \caption{Focal} 
    \vspace{1ex}
  \end{minipage} 
  \caption{The solid lines represent the differential cross sections in the azimuthal angle between the two hardest jets, $\Delta \Phi$, for p-p
 and p-Pb collisions computed using \KaTie\ and the ITMD approach. The error bands represents uncertainty due to scale variation from $(p_{T1}+p_{T2})/2$ by a factor of 1/2 and 2. The dotted lines represent the differential cross sections taking into account the non-perturbative correction factors from {\Pythia}. Similarly, the lower band represent uncertainty due to scale variation multiplied by the correction factor, taking into account statistical errors from \Pythia. The top two and the bottom left plots correspond to the FCal ATLAS kinematics, the bottom right plot corresponds to the FoCal upgrade of the ALICE.}
  \label{fig:katie_bands} 
\end{figure}

The partonic cross sections were calculated using the \KaTie\ Monte Carlo program~\cite{vanHameren:2016kkz} within the ITMD factorization scheme introduced above, for the following set of cuts on the transverse momenta $p_{T1},p_{T2}$ of the two leading jets: 
\begin{enumerate}[label={\it\roman*}$\,$)]%,leftmargin=35pt]
    \item $28\, \mathrm{GeV} < p_{T1},p_{T2} < 35\, \mathrm{GeV}$,
    \item $35\, \mathrm{GeV} < p_{T1},p_{T2} < 45\, \mathrm{GeV}$,
    \item $35\, \mathrm{GeV} < p_{T1} < 45\, \mathrm{GeV}$ and  $28\, \mathrm{GeV} < p_{T2} < 35\, \mathrm{GeV}$,
    \item $p_{T1}, p_{T2} > 10\, \mathrm{GeV}$.
\end{enumerate}
The first three cuts are tailored to the FCal calorimeter of the ATLAS detector for which  jets were considered in the rapidity range $2.7<y^{\star}_{1},y^{\star}_{2}<4.0$ in both the proton-proton and the proton-nucleon center of mass frame. The last set of cuts is for the planned ALICE upgrade FoCal and are used in the rapidity range $3.8<y^{\star}_{1},y^{\star}_{2}<5.1$. The positive rapidity windows correspond to a positive $z$-component for the proton momentum in p-Pb collisions. The jets were defined using the anti-$k_T$ jet clustering algorithm \cite{Cacciari:2008ab} with a radius parameter $R = 0.4$. The factorization and renormalization scales have been set to $(p_{T1}+p_{T2})/2$. The shaded bands in Fig.~\ref{fig:katie_bands} represent the error due to the variation of this value by a factor of 1/2 and 2.

In our computation within the ITMD framework, we included the following partonic channels, for five quark flavors:
\begin{equation}
    qg^*\longrightarrow qg, \quad \quad gg^* \longrightarrow gg\,,
\end{equation}
where the $*$ represents the off-shell gluon. The channel $ gg^*\longrightarrow \overline q q$ was neglected as the contribution of this channel is
small in considered kinematics \cite{Kutak:2012rf,vanHameren:2016ftb}. The  gluon distributions necessary for the ITMD framework calculated in \cite{vanHameren:2016ftb} were based on the Kutak-Sapeta (KS) fit of the dipole gluon density \cite{Kutak:2012rf}.
For the collinear PDFs in the ITMD framework, we used CTEQ10NLO set \cite{Lai_2010} from LHAPDF6 \cite{Buckley_2015}. 

The cross sections computed in the ITMD framework are obtained at the parton level. In order to estimate the effects due to the final state shower as well as  hadronization, we use the \Pythia\ Monte Carlo event generator \cite{pythia:2006ab, pythia:2015ac} version 8.307 with the default tunes. We used the NNPDF23NLO set \cite{Ball:2013ab} to describe the proton structure, and nCTEQ15WZ set \cite{kusina:2020ab} for the nuclear PDF necessary for the simulatuon of p-Pb collisions. 
The detailed procedure as follows:
\begin{enumerate}
    \item we simulate p-p collisions at parton level with the Initial State (IS) shower only using \Pythia; such setup is supposed to include similar physics to the \KaTie\ simulation, because the TMD gluon distributions mimic the IS showers,
    \item we turn on the Final State (FS) shower, hadronization and MPI effects; by comparing this with the previous calculation we estimate the correction factor,
    \item we apply similar procedure to p-Pb process, 
    \item we superimpose the correction factors to the \KaTie\ results, to obtain hadron-level cross sections.
    %with NLO collinear parton densities. 
    %where the factorization and renormalization scales were set to be the smaller of the squared transverse masses of the two outgoing particles.
\end{enumerate}

\begin{figure} 
  \begin{minipage}[b]{0.5\linewidth}
    \centering
    \includegraphics[width=1\linewidth]{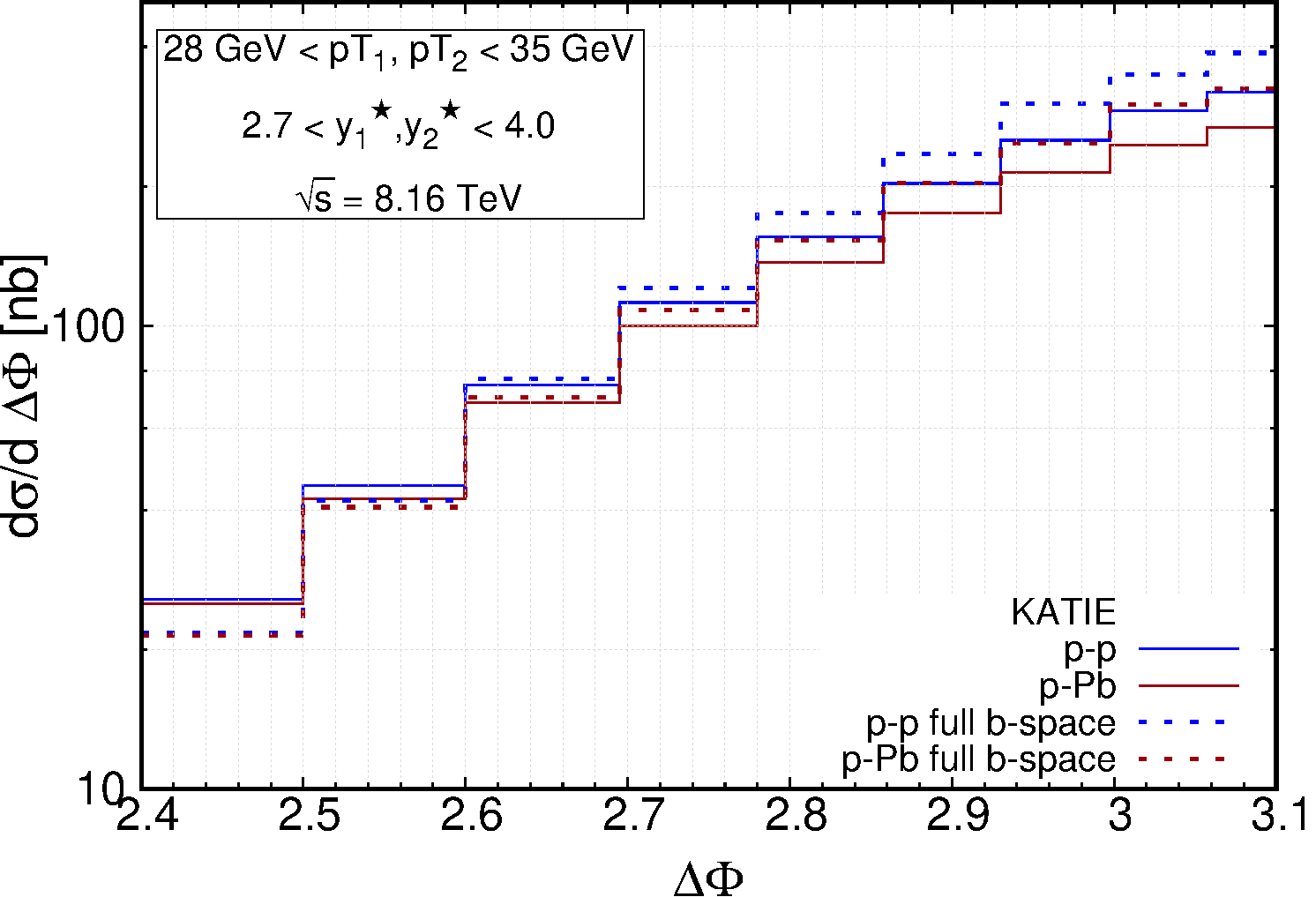} 
   %\caption{(a)} 
    \vspace{1ex}
  \end{minipage}%%
  \begin{minipage}[b]{0.5\linewidth}
    \centering
    \includegraphics[width=1\linewidth]{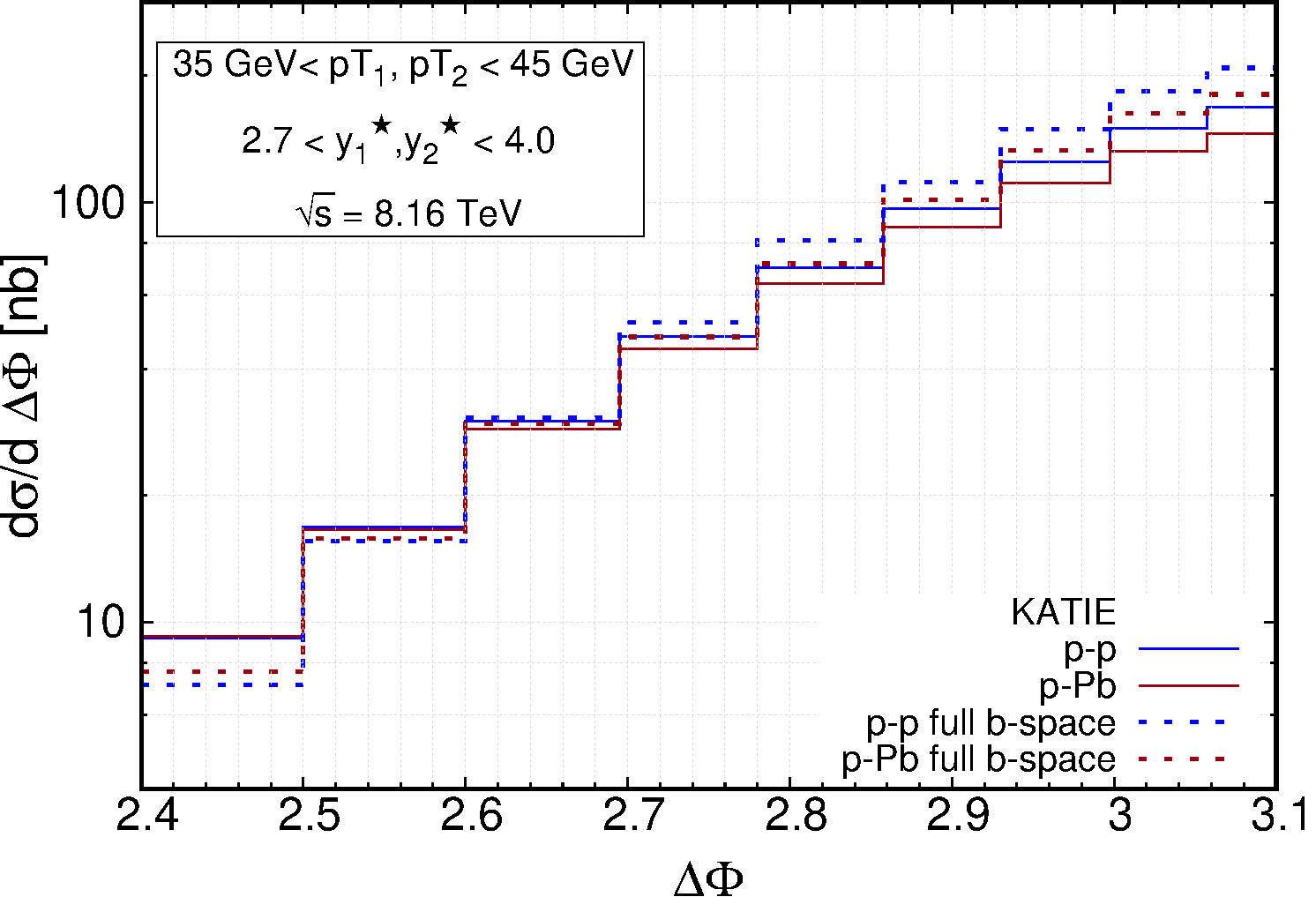} 
    %\caption{At2} 
    \vspace{1ex}
  \end{minipage} 
  \begin{minipage}[b]{0.5\linewidth}
    \centering
    \includegraphics[width=1\linewidth]{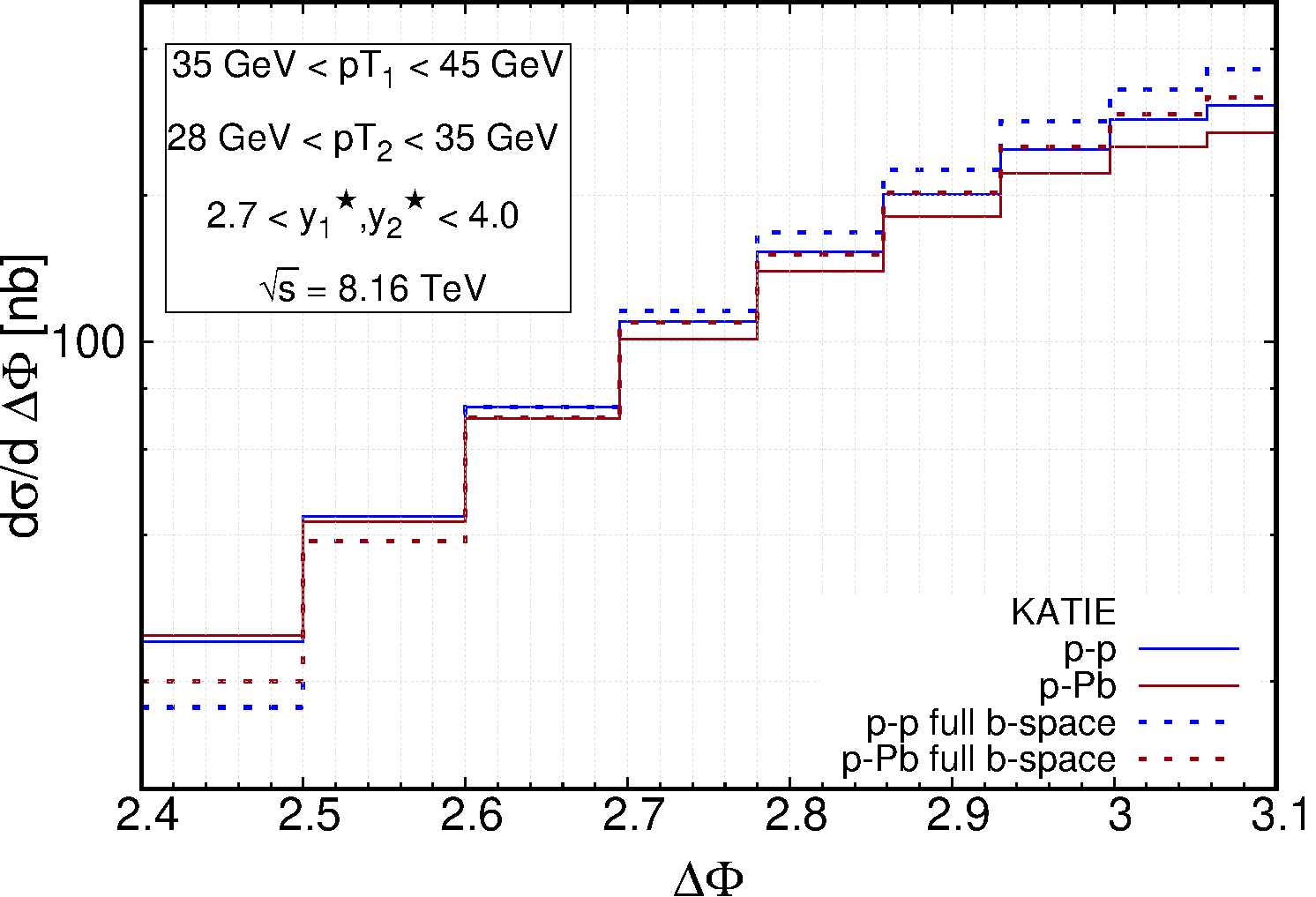} 
   % \caption{AT3} 
    \vspace{1ex}
  \end{minipage}%% 
  \begin{minipage}[b]{0.5\linewidth}
    \centering
    \includegraphics[width=1\linewidth]{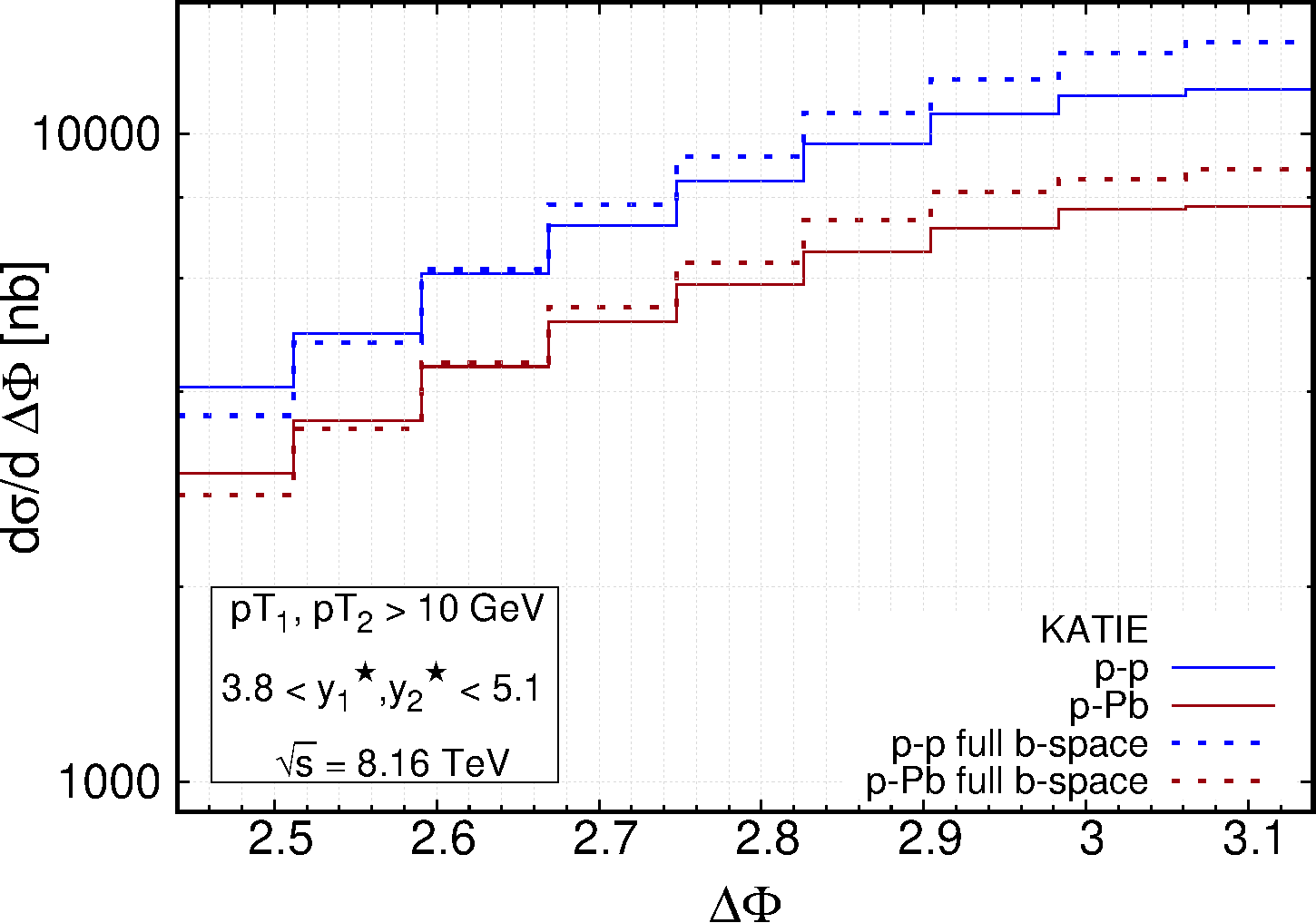} 
   % \caption{Focal} 
    \vspace{1ex}
  \end{minipage} 
  \caption{The differential cross sections in the azimuthal angle between the two hardest jets, $\Delta \Phi$, for p-p
 and p-Pb collisions computed from \KaTie\  using the ITMD factorisation formula with: the simplified Sudakov resummation Eq.~(\ref{eq:itmd}) (solid lines), the full $b$-space resummation Eq.~(\ref{eq:itmd_Sud}) (dotted lines). The top two and the bottom left plots correspond to FCal ATLAS kinematics, while the bottom right plot corresponds to the FoCal upgrade of ALICE.}
  \label{fig:only_katie} 
\end{figure}

In Fig.~\ref{fig:only_katie} we show the results  of the calculations, both for the ATLAS and ALICE kinematic region, using the ITMD factorisation formula with the Sudakov resummation in the simplified scheme of Eq.~(\ref{eq:itmd}), as compared to calculations based on the full $b$-space resummation Eq.~(\ref{eq:itmd_Sud}).
The calculations are  done  both for  p-p and p-Pb systems. We see that, overall, the results are similar.
However, the full $b$-space resummation gives slightly less decorrelation than the factorized approach. Within the accuracy of our LO predictions, both approaches can be treated on equal footing.
This is evident from Fig.~\ref{fig:katie_bands}, where we see that within the uncertainties obtained by varying the factorization/renormalization scales, the difference between full $b$-space and the factorized Sudakov form factor washes out. 
In addition, the difference between both Sudakov resummation schemes cancels to large extent in the ratio of p-Pb and p-p cross sections.

\begin{figure} 
  \begin{minipage}[b]{0.5\linewidth}
    \centering
    \includegraphics[width=1\linewidth]{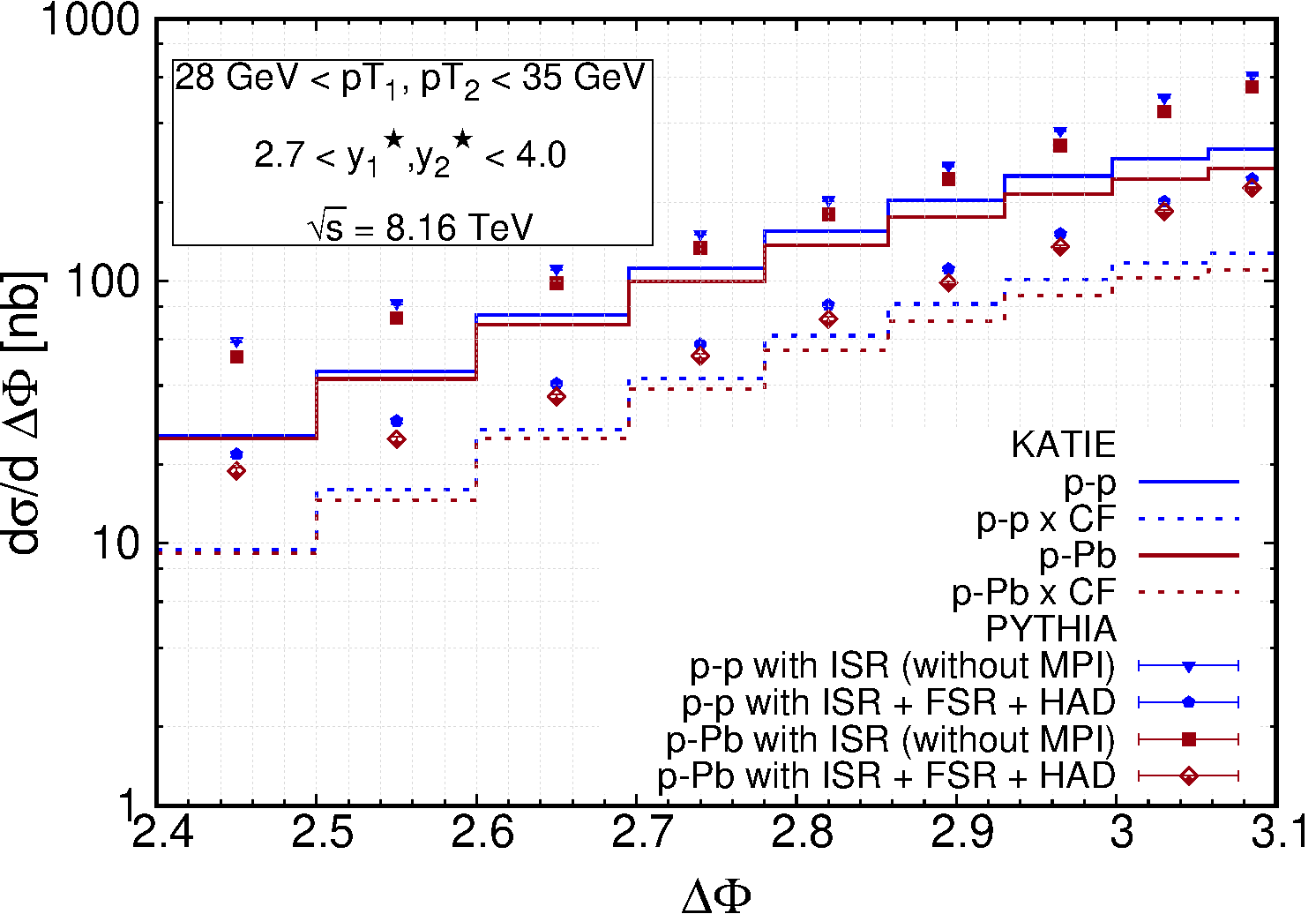} 
   %\caption{(a)} 
    \vspace{1ex}
  \end{minipage}%%
  \begin{minipage}[b]{0.5\linewidth}
    \centering
    \includegraphics[width=1\linewidth]{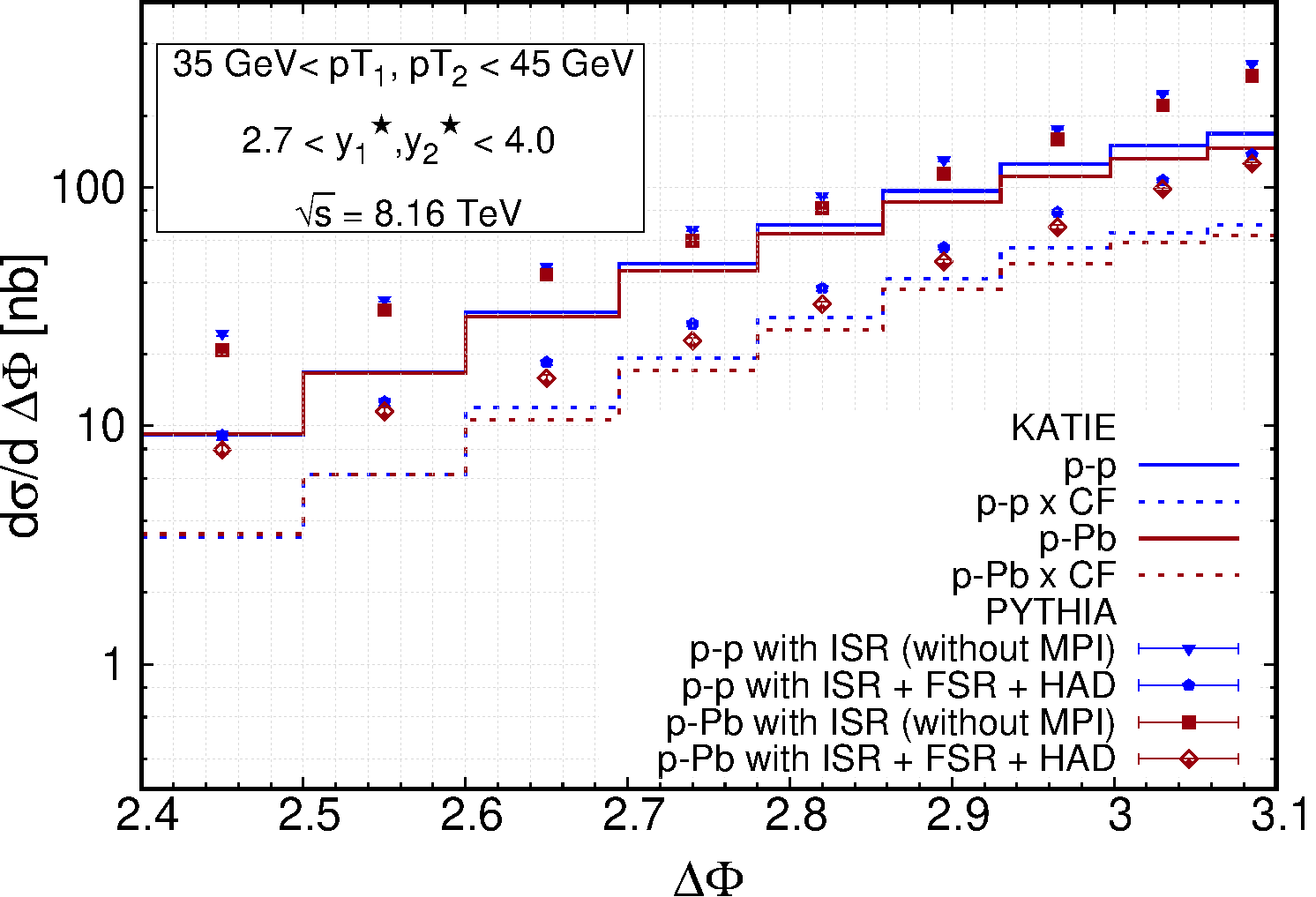} 
    %\caption{At2} 
    \vspace{1ex}
  \end{minipage} 
  \begin{minipage}[b]{0.5\linewidth}
    \centering
    \includegraphics[width=1\linewidth]{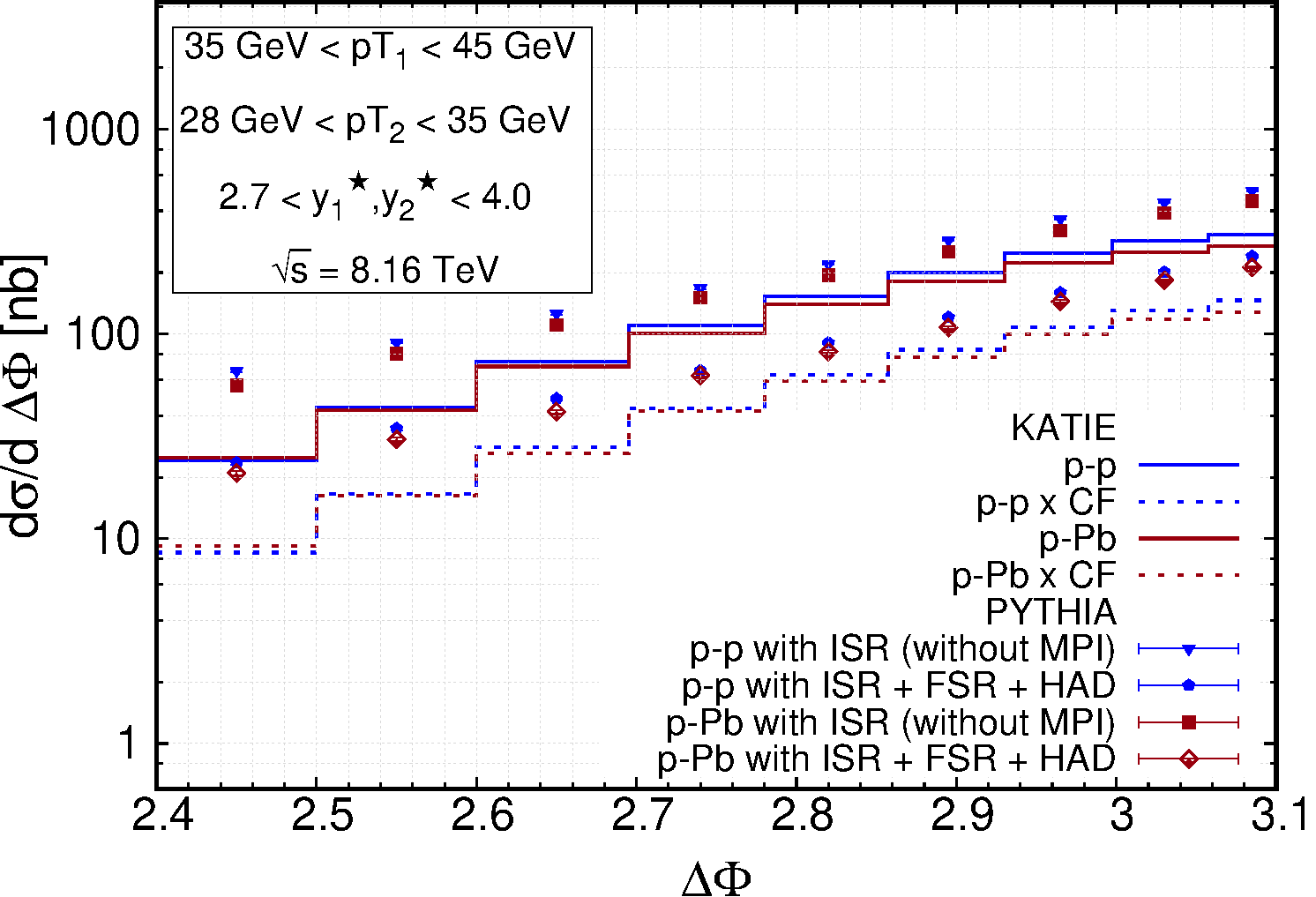} 
   % \caption{AT3} 
    \vspace{1ex}
  \end{minipage}%% 
  \begin{minipage}[b]{0.5\linewidth}
    \centering
    \includegraphics[width=1\linewidth]{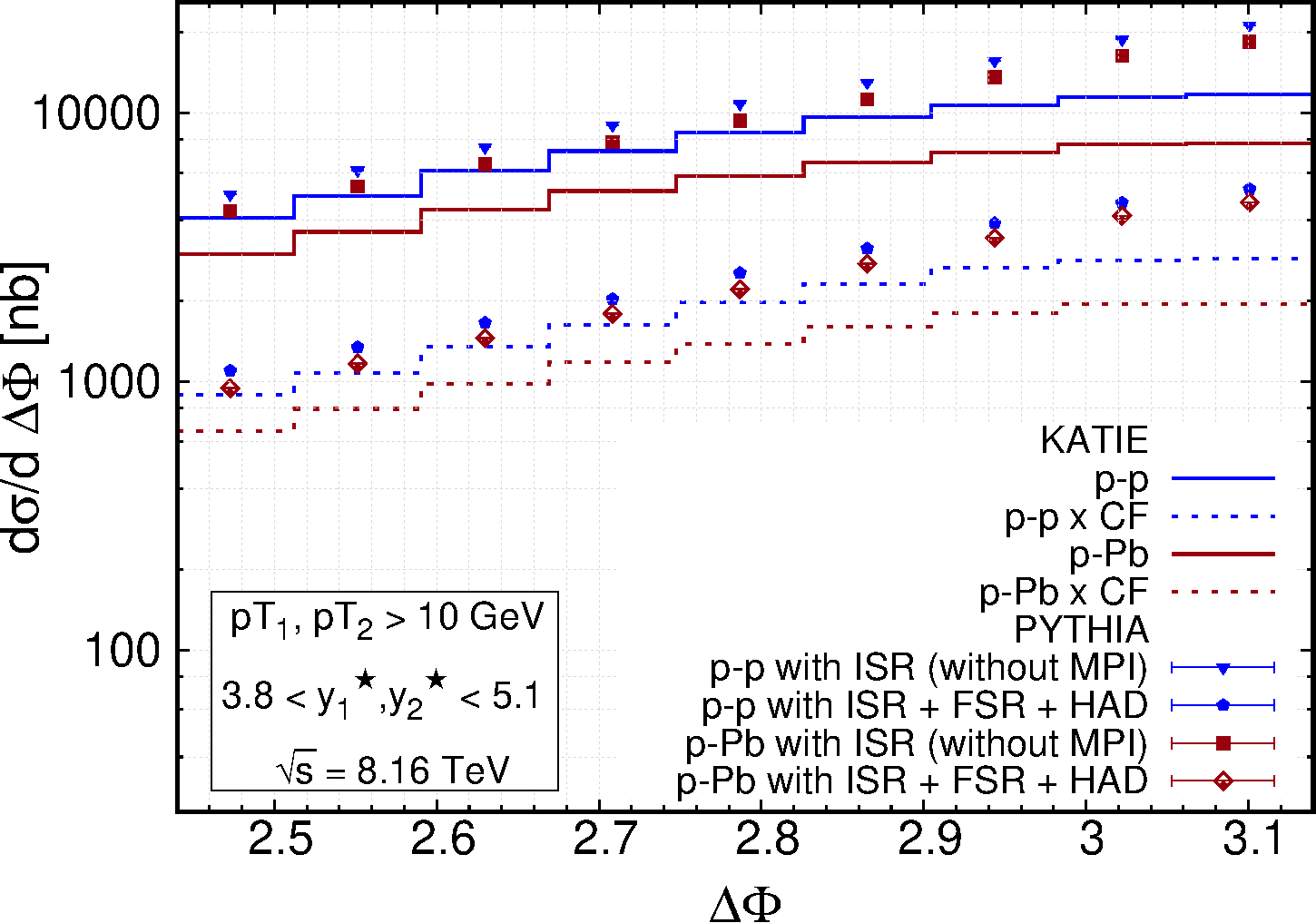} 
   % \caption{Focal} 
    \vspace{1ex}
  \end{minipage} 
  \caption{The differential cross sections in the azimuthal angle between the two hardest jets, $\Delta \Phi$, for p-p
 and p-Pb collisions computed using \KaTie\ with the ITMD approach (solid lines), {\Pythia} with various components (points) and the \KaTie\ with the non-perturbative correction factor extracted from {\Pythia} (dotted lines). The top two and the bottom left plots correspond to the FCal ATLAS kinematics, the bottom right plot corresponds to the FoCal upgrade of the ALICE.}
  \label{fig:katie_pythia} 
\end{figure}

In Fig.~\ref{fig:katie_pythia} we compare results obtained  within the ITMD approach to {\Pythia} calculations. We use only the factorized Sudakov resummation for simplicity. We observe that parton-level  {\Pythia} results, with only the initial-state shower applied, are above the ITMD results, and attribute this to the difference between linear and nonlinear evolutions.
One can also see that the ITMD results for the p-p and p-Pb spectra approach each other at small $\Delta\Phi$, while the  {\Pythia} results are shifted by a constant value for all values of $\Delta\Phi$. The behavior of the ITMD result is an expected manifestation of saturation effects. They are larger at large $\Delta\Phi$, leading to a more pronounced difference between the p-p and p-Pb curves at larger values of $\Delta\Phi$.
The final state shower, hadronization and MPI, essentially decrease the cross section, not changing the distribution shape too much, especially for larger transverse momenta. 

The extracted correction factors are applied to the \KaTie\ results, see Fig.~\ref{fig:katie_bands} (see also comparison with \Pythia\ results in Fig.~\ref{fig:katie_pythia}). The error bands for calculations with the correction factor are combinations of the scale variation error and the statistical error from \Pythia.  

\begin{figure} 
  \begin{minipage}[b]{0.5\linewidth}
    \centering
    \includegraphics[width=1\linewidth]{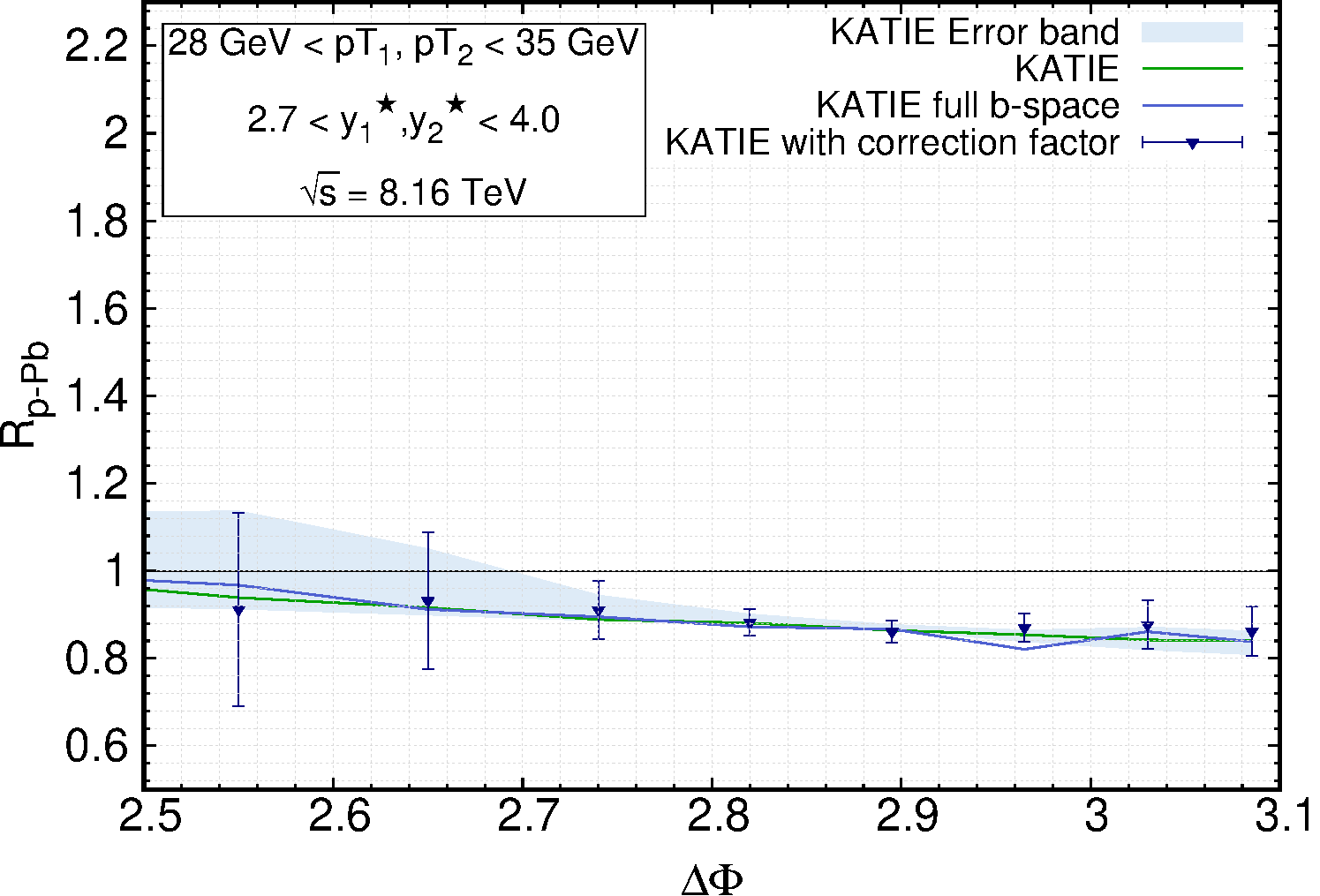} 
   %\caption{(a)} 
    \vspace{1ex}
  \end{minipage}%%
  \begin{minipage}[b]{0.5\linewidth}
    \centering
    \includegraphics[width=1\linewidth]{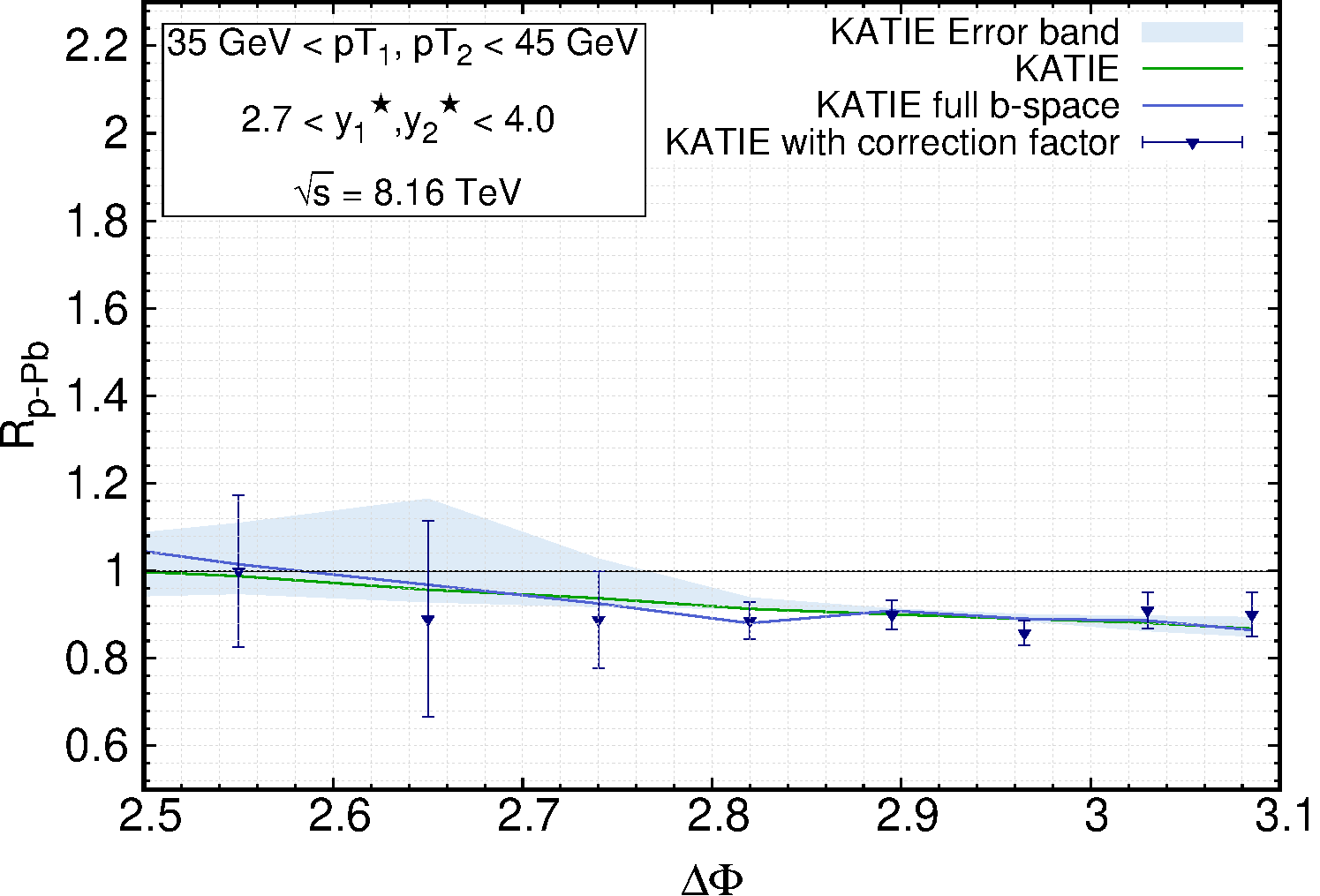} 
    %\caption{At2} 
    \vspace{1ex}
  \end{minipage} 
  \begin{minipage}[b]{0.5\linewidth}
    \centering
    \includegraphics[width=1\linewidth]{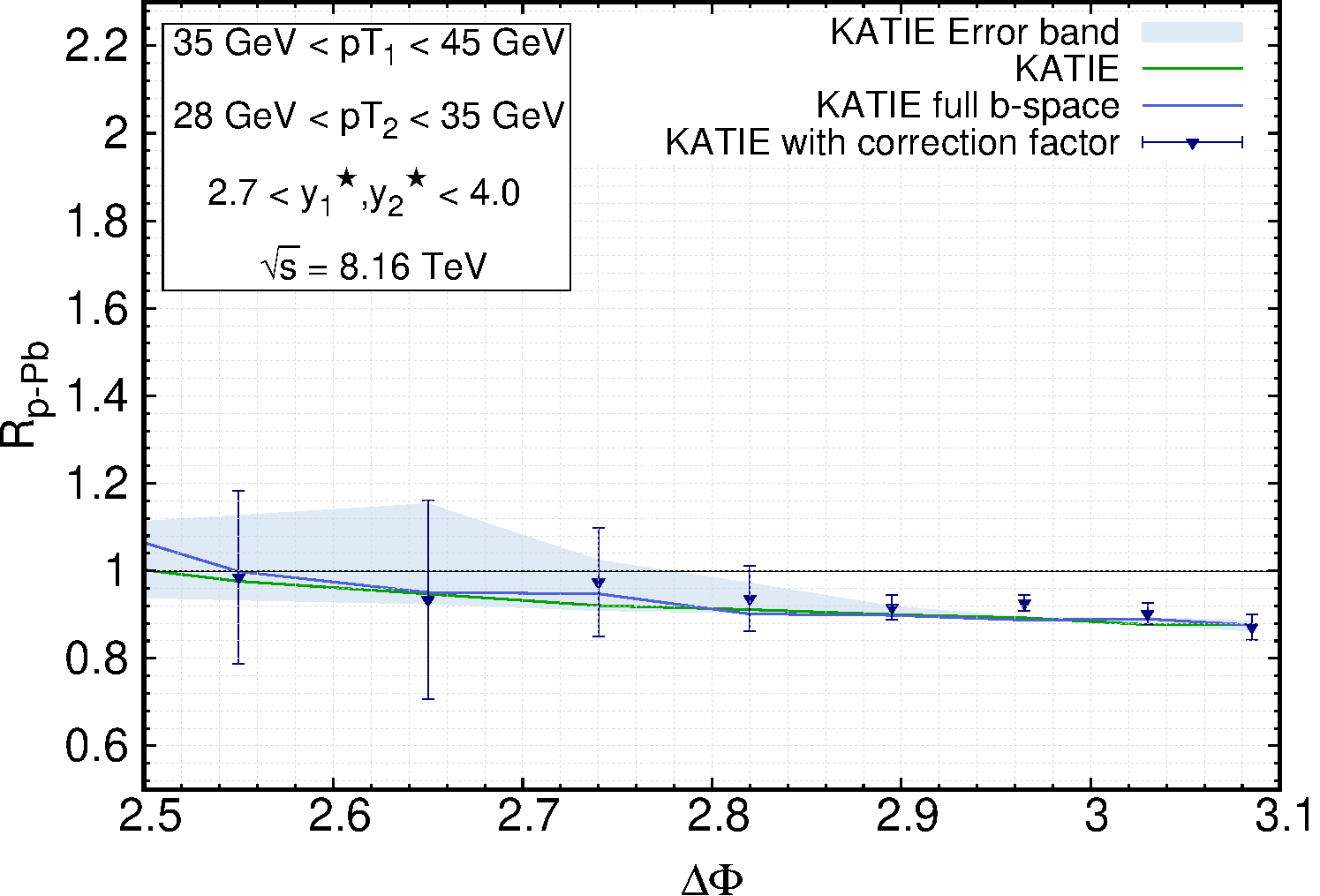} 
   % \caption{AT3} 
    \vspace{1ex}
  \end{minipage}%% 
  \begin{minipage}[b]{0.5\linewidth}
    \centering
    \includegraphics[width=1\linewidth]{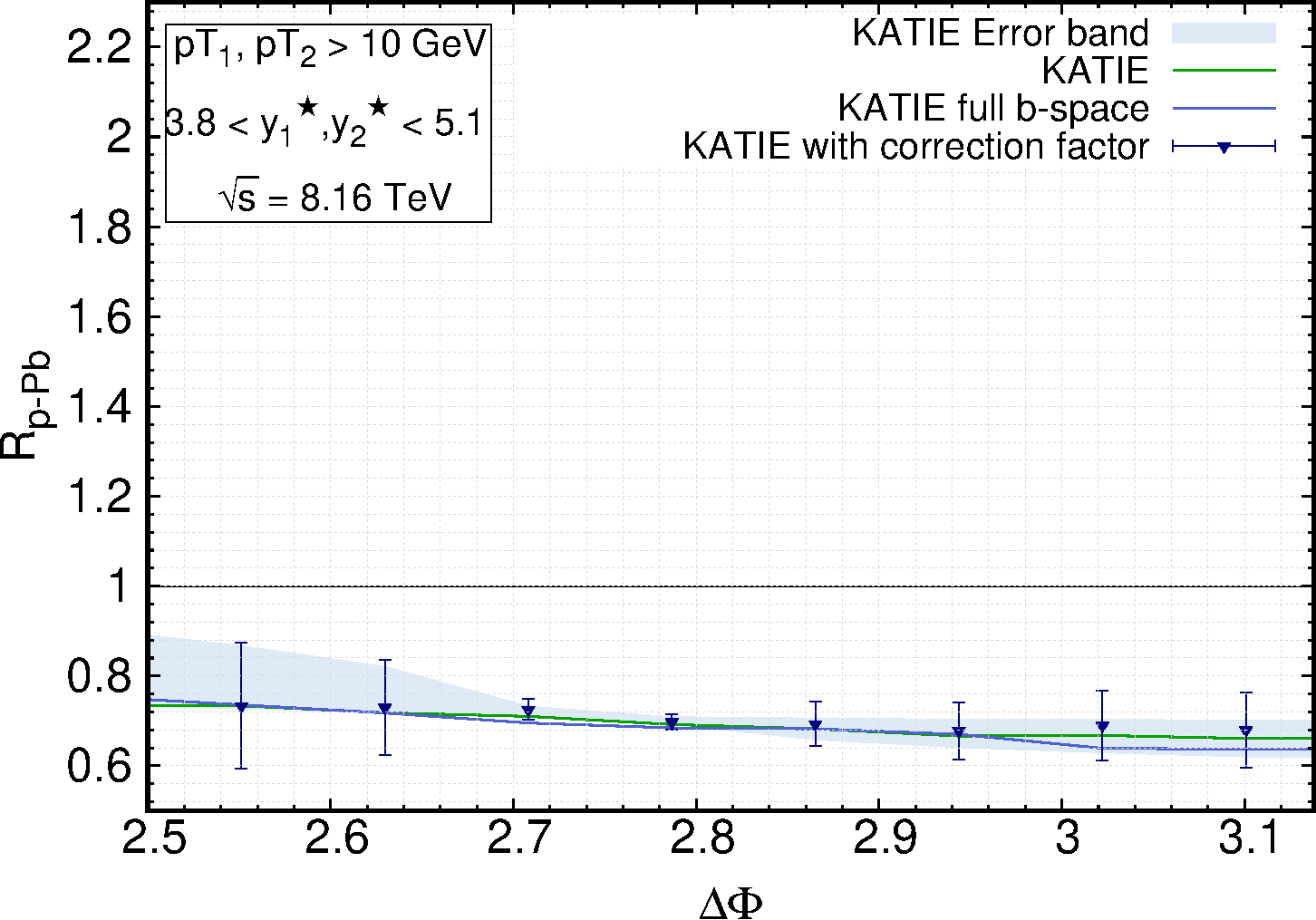} 
   % \caption{Focal} 
    \vspace{1ex}
  \end{minipage} 
  \caption{Nuclear modification ratio $R_{\mathrm{p-Pb}}$ as a function of the azimuthal angle between the jets $\Delta \Phi$ for the FCal ATLAS and ALICE FoCal kinematics. The error bands represents uncertainty associated with the \KaTie\ ITMD results due to the variation of the factorization scale from $(p_{T1}+p_{T2})/2$ by a factor of 1/2 and 2. The $\triangledown$ points represent $R_{\mathrm{p-Pb}}$ obtained from \KaTie\ multiplied by the non-perturbative correction factors from {\Pythia}. The error bars associated with these account for both the error in \KaTie\ via the variation of factorization scale plus the statistical uncertainties associated with the correction factors.}
  \label{fig:R_PA} 
\end{figure}

In Fig.~\ref{fig:R_PA} we  show the results for nuclear modification ratio $R_{pA}$ which determines the strength of suppression due to the saturation effects as one goes from a proton to a nuclear target. First of all we see that the suppression is quite large, about 20$\%$, for the FoCal upgrade of ALICE. The saturation signal persists even after including the correction due to the hadronization and other effects, which is an important result.
The second important observation is that the difference between the full $b$-space Sudakov resummation and the simplified approach cancels out to large degree in the nuclear modification ratio. Thus, the saturation signal is not much affected by the details of the Sudakov suppression of the back-to-back peak. For ATLAS kinematics, 
which is restricted to a slightly more central region,
we see  similar trends as for ALICE kinematics but the suppression due to saturation is smaller.  

%-----------------------------------------------------------------------
\section{Summary}
\label{sec:Summary}
We provided state-of-the-art predictions for the cross-sections and the nuclear modification ratio $R_{pA}$ for di-jet production in forward-forward jets in kinematic domains covered by the FCal ATLAS detector and the planned FoCal upgrade of the ALICE experiment. 
The calculation is based on application of the ITMD factorization approach implementing the saturation and the kinematic twist corrections, together with the Sudakov resummation necessary for realistic description of azimuthal observables in jet production processes.
The Sudakov form factor was implemented using two approaches, a simplified approach, where the collinear PDF describing the dilute projectile is factorized and the full $b$-space resummation affecting also the latter. Both  frameworks give results that are close to each other for the considered kinematics with the ITMD result being below the  \Pythia\ result.
As the ITMD calculation is a parton level calculation, we used \Pythia\ in order to estimate corrections for hadronization, FSR shower and MPI effects. For nuclear targets, we used the nuclear PDFs in \Pythia\ to simulate the correction factor. The correction factor is not much sensitive to the actual PDF used. Therefore using the correction factor extracted from nuclear PDFs that do not -- at least explicitly -- have saturation effects on the top of the ITMD saturation framework is  a rough but realistic estimate.
We conclude that, taking into the account all the uncertainties, the measurement of the nuclear modification ratio will allow to determine the suppression due to saturation effects.

\section{Acknowledgements}
\label{sec:acknowledgements}
We would like to thank Jacek Otwinowski, Anne Sickless, Peter Steinberg, Marco Van Leeuven for stimulating discussions and Riccardo Longo, Farid Salazar and Bowen Xiao for useful correspondence. \\
KK  acknowledges the hospitality of the Nuclear Theory group at the BNL where part of the  project was realized. \\
This work is supported partially by grant V03319N from the common FWO-PAS exchange program. 
MAA is supported by Science, Technology \& Innovation Funding Authority (STDF), Egypt under Project ID: 30163.
AvH acknowledges the Polish National Science Centre grant no.\ 2019/35/B/ST2/03531. 
HK is supported by the National Science Centre, Poland grant no. 2021/41/N/ST2/02956. 
PK is supported by the Polish National Science Centre, grant no. DEC-2020/39/O/ST2/03011.
KK acknowledges the Polish National Science Centre grant no.\ DEC-2017/27/B/ST2/01985.

\bibliographystyle{unsrt} % sorts in the order of apperance
%\bibliography{sudakov}

\end{document}